# Epitaxial growth of boron carbide on 4H-SiC

BENAMRA Yamina[1,a*], AUVRAY Laurent[1,b], ANDRIEUX Jérôme[1,c], CAUWET François[1,d], ALEGRE Maria-Paz[2,e], LLORET Laurent[2,f], Daniel Araujo[2,g], Marina Gutierrez[2,h], FERRO Gabriel[1,i*],

[1]Laboratoire des Multimatériaux et Interfaces, Université de Lyon, 6 rue Victor Grignard, 69622, Villeurbanne, France

[2]Materials Science Department, Universidad de Cádiz, 11510 Puerto Real, Spain

[a]yamina.benamra@univ-lyon1.fr, [b]laurent.auvray@univ-lyon1.fr, [c]jerome.andrieux@univ-lyon1.fr, [d]francois.cauwet@univ-lyon1.fr, [e]maripaz.alegre@uca.es, [f]fernando.lloret@uca.es, [g]daniel.araujo@uca.es, [h]marina.gutierrez@uca.es, [i]ferro.gabriel@univ-lyon1.fr

**Abstract.** In this work, the successful heteroepitaxial growth of boron carbide ($B_xC$) on 4H-SiC(0001) 4° off substrate using chemical vapor deposition (CVD) is reported. Towards this end, a two-step procedure was developed, involving the 4H-SiC substrate boridation under $BCl_3$ precursor at 1200°C, followed by conventional CVD under $BCl_3$ + $C_3H_8$ at 1600°C. Such a procedure allowed obtaining reproducibly monocrystalline (0001) oriented films of $B_xC$ with a step flow morphology at a growth rate of 1.9 µm/h. Without the boridation step, the layers are systematically polycrystalline. The study of the epitaxial growth mechanism shows that a monocrystalline $B_xC$ layer is formed after boridation but covered with a B- and Si-containing amorphous layer. Upon heating up to 1600°C, under pure $H_2$ atmosphere, the amorphous layer was converted into epitaxial $B_xC$ and transient surface $SiB_x$ and Si crystallites. These crystallites disappear upon CVD growth.

## Introduction

Despite the recent significant development of 4H-SiC devices for high power - high voltage applications, this technology is only composed of SiC as the active semiconductor. Some attempts to integrate other semiconductors into 4H-SiC-based heterojunctions [1–3] have been made, but the performances of the obtained devices were somewhat deceiving, essentially due to the insufficient crystalline quality of the hetero-material. Boron carbide ($B_xC$), a compound that is better known for technical ceramic applications, is also a wide bandgap material (1.6 - 2.2 eV [4]) that could be highly compatible with SiC in terms of thermal and chemical stability. However, its electronic properties are very little known [4–8], probably due to the difficulty in elaborating single crystalline $B_xC$ either in bulk [9,10] or thin film form [11–16]. The determination of these properties requires, thus, first succeeding in growing a monocrystalline material such as a thin film on 4H-SiC substrate. Recently, very thin $B_xC$ films (20 nm) were successfully heteroepitaxially grown on 4H-SiC using unconventional techniques (thermal vapor deposition and pulsed laser deposition) [17]. A more classical and versatile technique to grow such layers, such as chemical vapor deposition (CVD), would be preferred for application purposes. But $B_xC$ heteroepitaxy using CVD has never been demonstrated so far. This is the goal of the present study.

## Experimental

CVD growth was performed in a homemade vertical cold wall reactor working at atmospheric pressure. Depositions were made on commercial 4° off-axis 4H-SiC(0001) Si face substrates (from SK Siltron) which were cleaned in a methanol ultrasonic bath before their introduction into the reactor. High-purity boron trichloride ($BCl_3$, 1% diluted in Ar) and propane ($C_3H_8$, 5% diluted in $H_2$) were used as boron and carbon precursors, respectively. 16 slm of high-purity dihydrogen was used as carrier gas. The sample holder, a SiC-coated graphite susceptor, was induction-heated, and its temperature was measured and monitored via an optical pyrometer.

After an in-situ removal of the substrate native oxide under $H_2$ at 1000°C for 5 min, epitaxial growth of $B_xC$ was obtained using a two-step procedure. First, a boridation step was applied, consisting of heating at 1200°C for 10 minutes under 2.5 sccm of $BCl_3$. The goal was to form a buffer layer of $B_xC$ on the 4H-SiC substrate, similarly to the carbonization step when growing 3C-SiC on silicon substrate [18]. Then, the temperature was increased with a ramp rate of 10°C/s up to 1600°C under $H_2$ only, at which point 2.5 sccm $BCl_3$ and 0.84 sccm $C_3H_8$ were added to start the $B_xC$ deposition with a C/B ratio in the gas phase equal to 1. To better assess the benefits of the boridation step, $B_xC$ layers were also grown without boridation, i.e., with a direct deposition at 1600°C, with the same precursor flows.

**Results and discussion**

*Demonstration of epitaxy*

The two-step growth procedure led to a homogeneous covering of the substrate with a film composed of triangular features all oriented in the same direction (Fig. 1a and 1b). The surface displays a rough but oriented step and terrace morphology, with an average root mean square (RMS) roughness of ~30 nm for a 20 x 20 µm² scan, as measured by atomic force microscopy (AFM). If the boridation step is skipped and the layer grown directly at 1600°C, the surface morphology is drastically different: the substrate is partially covered with crystallites of 10 to 30 µm lateral size (Fig. 1c).

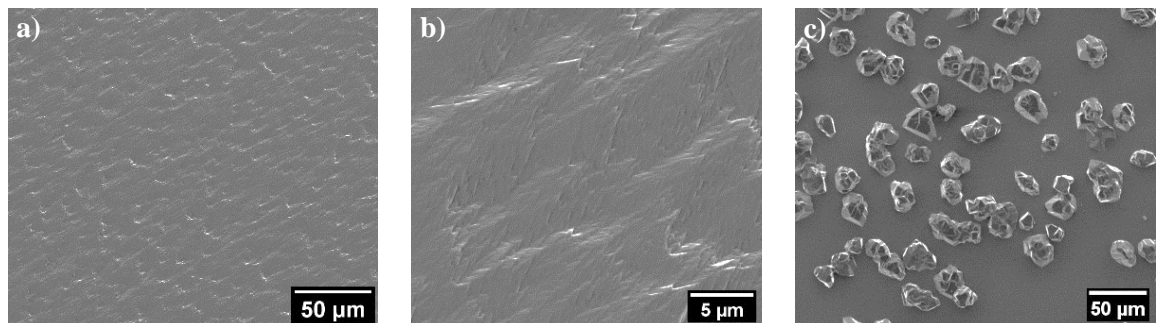

**Figure 1:** SEM images showing typical morphology of $B_xC$ films obtained on 4H-SiC using a) and b) the two-step growth procedure and c) a direct deposition at 1600°C.

X-ray diffraction (XRD) characterizations performed on these different layers are shown in Fig. 2. The sample grown with the boridation step gives a pattern composed of a single peak (besides the one of the substrate) attributed to the third reflection of the (0001) plane of boron carbide. The $B_xC$ layer seems highly oriented towards the same (0001) direction as the substrate. On the other hand, for the sample grown without boridation, several peaks are observed in the XRD pattern, which can be attributed to different crystalline orientations of a $B_xC$ phase. The crystallites formed under these conditions are obviously not epitaxial.

The epitaxial nature of the deposit grown using the boridation was confirmed by transmission electron microscopy (TEM) analyses. The cross-section image in Fig. 3 shows a defected but monocrystalline $B_xC$ layer of 1.9 µm on top of monocrystalline 4H-SiC. The layer is (0001) oriented, just like the substrate, which confirms that the heteroepitaxial growth was successful. The transition from SiC to $B_xC$ looks abrupt though a SiC "bump" can be seen at the interface (bottom right). A high density of crystalline defects is generated at the interface, probably due to the high lattice mismatch of 40% between $B_xC$ and 4H-SiC [19]. Interestingly, this defect density decreases significantly with increasing thickness. It should be noted that, despite the relatively high coefficient of thermal expansion (CTE) mismatch (~15%) between $B_xC$ and 4H-SiC, we did not observe the presence of cracks inside the epilayer.

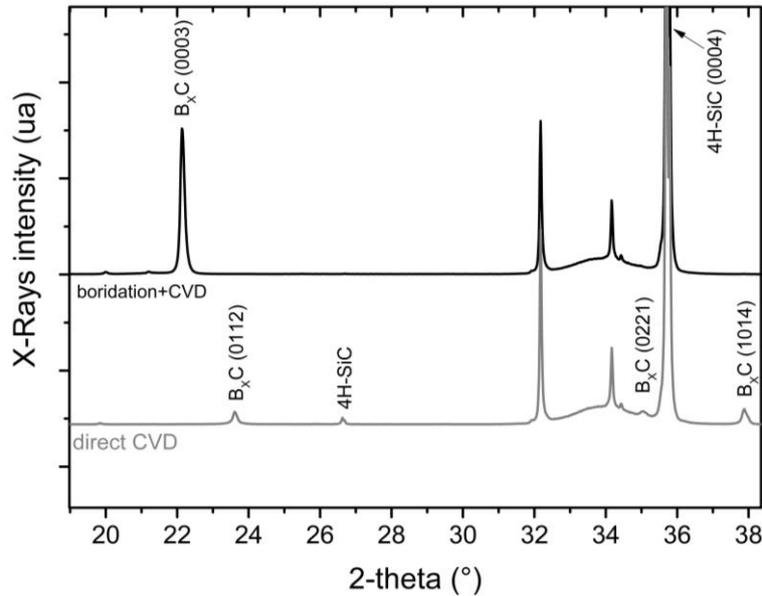

**Figure 2:** XRD patterns recorded on the same samples as shown in Fig. 1. Non-indexed peaks correspond to the apparatus's W and $K_\beta$ spectral lines.

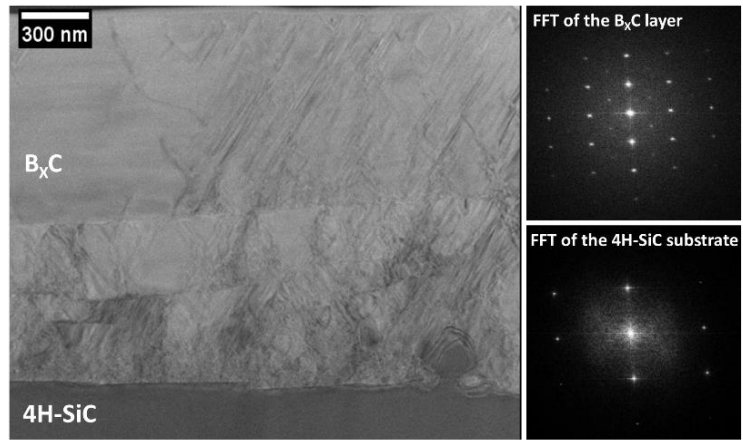

**Figure 3:** Low magnification cross-section TEM micrograph and corresponding diffraction patterns associated with the substrate (bottom) and the $B_xC$ epilayer (top).

*Role of the boridation step*

As evidenced by the previous results (Fig. 1 and 2), boridation is essential for obtaining heteroepitaxial growth of $B_xC$. To better understand the role of this step, a growth run was stopped just after the boridation step. SEM observation of the resulting sample showed a very smooth and homogeneous surface (Fig. 4a), with a measured RMS roughness as low as ~0.7 nm for a 10 x 10 µm² AFM scan. Cross-section TEM characterization of this sample revealed the formation of a bi-layer deposit composed of a 155 nm thick B- and Si-containing amorphous layer on top of a 35 nm thick highly twinned but monocrystalline $B_xC$ layer (Fig. 4b). This $B_xC$ layer is in direct contact with the 4H-SiC substrate and the interface is abrupt. Therefore, it can be concluded that the boridation works similarly to the carbonization step in the 3C-SiC/Si heteroepitaxy, excepting the presence of excess B-Si film on top of the $B_xC$ buffer layer. Nevertheless, this amorphous layer does not prevent further $B_xC$ epitaxy from occurring during CVD.

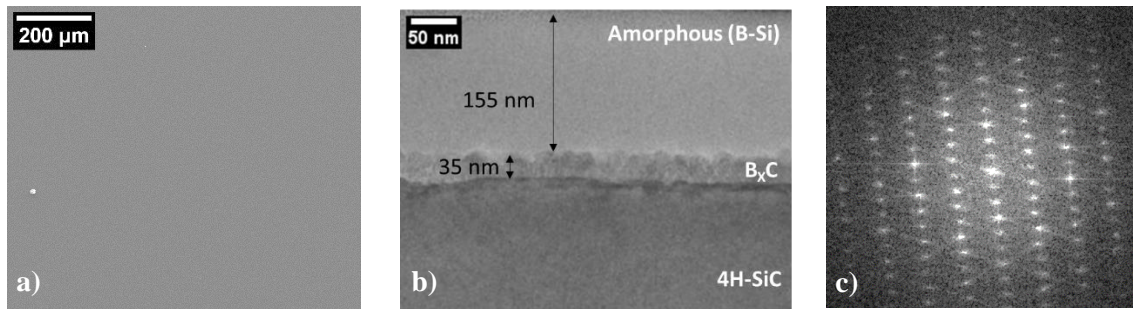

**Figure 4:** a) SEM image of the surface of a sample after the boridation step only and b) low magnification cross-section TEM micrograph of the same sample, and c) SAED pattern recorded on the B$_x$C monocrystalline buffer layer.

In order to better understand why the amorphous layer does not affect the epitaxial growth, we stopped an experiment just before the CVD growth, i.e., upon reaching 1600°C after the boridation step. The surface of the resulting sample was found to be composed of three different phases, which could be better distinguished using the Z-contrast in backscattered electron SEM imaging (Fig. 5a): i) light-grey facetted crystals of several µm lateral size, ii) small white crystallites of ~0.5 µm size and iii) a dark grey phase obviously covering the 4H-SiC surface completely. The white crystallites were identified as pure Si by Raman spectroscopy and energy-dispersive X-ray spectroscopy (EDS) (not shown). EDS analysis on the bigger light-grey crystals revealed a silicon boride phase containing ~19 at% Si. This is not that far from the theoretical composition of the SiB$_6$ phase (14.6 at% Si).

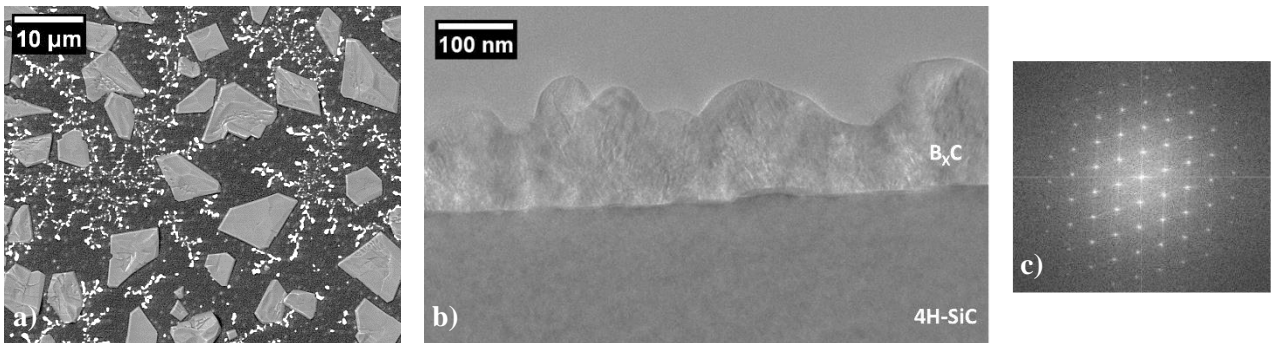

**Figure 5:** a) SEM backscattered electron image of the sample surface obtained by stopping the growth upon reaching 1600°C (after boridation), b) high magnification cross-section TEM micrograph of the same sample, and c) SAED pattern recorded on the B$_x$C monocrystalline layer.

TEM characterization of this sample (Fig. 5b and 5c) evidenced a thicker (~150 nm) and rougher monocrystalline layer of B$_x$C, which should correspond to the dark-grey phase in Fig. 5a. This thickening occurred during the 40 s heating from 1200°C up to 1600°C under H$_2$ only. As the SiC substrate is the only source of additional C atoms in the system during the heating ramp, this implies further reactivity of the substrate. This suggests that despite the presence of the B$_x$C buffer layer, the B-containing amorphous layer interacts with SiC, leading to the epitaxial thickening and a Si enrichment of the surface. Such interaction should imply some cross-diffusion of C, B, and Si atoms through the B$_x$C layer. The general (unbalanced) reaction could be written as follows:

$$SiC + B\text{-}Si \Rightarrow B_xC_{(epi)} + SiB_y + Si$$

Interestingly, the excess of B and Si at the surface (under the form of SiB$_6$ and elemental Si) does not seem to affect the epitaxial regrowth on top of the newly formed B$_x$C monocrystalline layer upon reaching 1600°C. Therefore, we can propose that when starting the CVD, the Cl atoms brought by the BCl$_3$ precursor could etch away this excess of Si while the excess of B could be converted into

$B_xC$ by reacting with the $C_3H_8$ precursor. When a stable $B_xC$ growth front is obtained, step flow growth can then occur, leading to the morphology shown in Fig. 1. The present explanation, which is far from complete, would need further investigation of the transition step and the early stage of CVD growth. This is in progress.

Finally, it is essential to state that the mechanism leading to $B_xC$ epitaxy on 4H-SiC is very robust despite its complexity. Our experience shows that the epitaxial process can be extended to a broader range of experimental conditions. Indeed, we also obtained $B_xC$ heteroepitaxy using different: i) temperatures and durations of boridation, ii) temperatures of CVD growth, iii) C/B ratios in the gas phase, or iv) polarity of the 4H-SiC substrates.

## Summary

For the first time, we showed successful heteroepitaxial boron carbide growth on 4H-SiC(0001) substrate by CVD. The boridation step appeared to be more complex than expected since it involved the formation of a B-Si amorphous layer on top of the monocrystalline $B_xC$ buffer layer. This B-containing layer does not prevent further epitaxy and even contributes to the initial thickening of the buffer layer. The transient formation of Si-containing phases does not affect the CVD regrowth, suggesting their rapid Cl-induced etching. The overall process is very robust and reproducible over a wide range of growth conditions.

## Acknowledgments

Electron microscopy was performed at the Centre Technologique des Microstructures of the University of Lyon (CTμ).